# The Upcoming Mutual Event Season for the Patroclus – Menoetius Trojan Binary

**W.M. Grundy[a], K.S. Noll[b], M.W. Buie[c], and H.F. Levison[c]**
    a. Lowell Observatory, Flagstaff AZ 86001.
    b. NASA Goddard Space Flight Center, Greenbelt MD 20771.
    c. Southwest Research Institute, Boulder CO 80302.

## Abstract

We present new Hubble Space Telescope and ground-based Keck observations and new Keplerian orbit solutions for the mutual orbit of binary Jupiter Trojan asteroid (617) Patroclus and Menoetius, targets of NASA's Lucy mission.  We predict event times for the upcoming mutual event season, which is anticipated to run from late 2017 through mid 2019.

## Introduction

As the first known binary Jupiter Trojan, Patroclus and Menoetius provided the first opportunity to directly measure the mass and thus bulk density of objects from this population. Marchis et al. (2005; 2006) determined their mutual orbit to have a period of 4.283 ± 0.004 days and a semimajor axis of 680 ± 20 km, leading to a system mass of (1.36 ± 0.11) x $10^{18}$ kg.  When the plane of their mutual orbit is aligned with the direction to the Sun or to an observer, Patroclus and Menoetius take turns eclipsing or occulting one another.  Such an alignment occurs during mutual event seasons, twice every ~12 year orbit around the Sun.  Spitzer Space Telescope thermal observations of Patroclus – Menoetius mutual events enabled Mueller et al. (2010) to determine that the surface thermal inertia is relatively low, consistent with mature regolith.  The thermal infrared observations also enabled them to better constrain the sizes of the objects, leading to a mean bulk density of 1.08 ± 0.33 g $cm^{-3}$, with the density uncertainty being dominated by the uncertainty in the volumes of the two bodies.

    The first spacecraft exploration of the Jupiter Trojans is to be done by NASA's Lucy probe, planned for launch in 2021.  After flybys of four objects in the L4 (leading) Trojan cloud, Lucy will fly past Patroclus and Menoetius in the L5 (trailing) cloud in 2033.  This pending visit motivates additional observational studies of the Patroclus – Menoetius system in support of detailed encounter planning.  An opportunity to refine the orbit as well as other properties of the system presents itself with a new mutual event season (see http://fredvachier.free.fr/binaries/phemu.info.php).  However, enough time has passed that the uncertainties associated with the Marchis et al. orbit solution are too large to permit precise predictions for the upcoming mutual events.  Additionally, it is unknown whether the orbit could be changing over time, due to some

as-yet unaccounted for perturbation.

## Observations

To obtain a new, independent mutual orbit for Patroclus and Menoetius ahead of the upcoming mutual event season, we observed the system at five epochs during May-June 2017, using the Wide Field Camera 3 (WFC3) ultraviolet and visible (UVIS) camera (e.g., Dressel et al. 2012) of the Hubble Space Telescope (HST). Our observing sequence for each of the five epochs consisted of 16 dithered frames, the majority of them using the broadband *F555W* filter (the number in the filter name indicates the central wavelength in nanometers). A few frames were acquired using *F438W* and *F814W* filters, resulting in color photometry that will be analyzed separately.

The system was observed again in December 2017, using the laser guide star adaptive optics (LGS AO) system at the 10 m Keck II telescope on Mauna Kea. Nine 60 second integrations were obtained with the NIRC2 narrow camera, using an H-band filter, at a mean time of 2017-Dec-09 14:36 UTC.

## Analysis

The positions of the system components in detector coordinates were obtained via familiar PSF-fitting methods using the Tiny Tim PSF model that captures many of the idiosyncrasies of the HST optical system (Krist and Hook 2004; Krist et al. 2011). We modeled two point sources to simultaneously fit for a pair of $x$, $y$ positions and fluxes, six free parameters in all. The nearly negligible sky background was removed before fitting by subtracting the mean of the sky.

The first step was to find a best-fit set of model parameters for each image using an amoeba downhill simplex method (Nelder and Mead 1965; Press et al. 1992). Given a starting location defined by the best-fit model, we used a Metropolis-Hasting Markov chain Monte Carlo (MCMC) sampling algorithm (e.g., Hastings 1970) to derive a probability density function (PDF) for each of the fitted values (*mcmcsamp.pro* from https://www.boulder.swri.edu/~buie/idl). This sampling was performed on all variables at the same time to capture the parameter uncertainty as well as any potential correlations between parameters. The best-fit values used the total counts for the flux while during the MCMC sampling, this variable was sampled in magnitude space. One important modification of this algorithm was to impose a domain restriction on the fitted parameters during the PDF sampling. For this work, there was no restriction on magnitude but the position was restricted to be within ±0.6 pixels of the initially fitted location.

These PDFs in pixel coordinates comprise the actual measurement but need to be converted to sky-plane positions for orbit fitting. There were insufficient on-chip astrometric reference stars so our measurements are strictly differential. The raw pixel location is converted to RA, Dec using the *xy2ad.pro* routine from the Astronomy User's Library (https://idlastro.gsfc.nasa.gov).

The Keck observations were reduced by fitting a pair of elliptical Lorentzian PSFs to each frame, as described by Grundy et al. (2015). Uncertainties in the relative astrometry were estimated from the dispersion of the measurements from the nine frames.

An additional data point was taken from the 2013 October 21 stellar occultation. Numerous ground-based observing stations yielded chords across the system, providing accurate size and shape information, as well as pinpointing the relative locations of the two bodies at that epoch (Buie et al. 2015). The resulting astrometric data set is listed in Table 1.

| Table 1 - Relative Astrometry | | | | | | |
|---|---|---|---|---|---|---|
| **Observation date and mid-time (UTC)** | **Observation type** | $r$ (AU) | $\Delta$ | $g$ (deg.) | $\Delta x$ | $\Delta y$ |
| | | | | | (arcsec) | |
| 2013-10-21 06:43 | Stellar occultation | 4.674 | 3.710 | 3.39 | −0.2463(30) | −0.0185(30) |
| 2017-05-20 13:09 | HST/WFC3 | 5.835 | 5.973 | 9.74 | +0.1434(19) | −0.0062(21) |
| 2017-05-29 22:44 | HST/WFC3 | 5.839 | 6.109 | 9.38 | −0.0194(28) | −0.0585(27) |
| 2017-06-08 00:27 | HST/WFC3 | 5.843 | 6.235 | 8.90 | −0.1138(21) | −0.0498(21) |
| 2017-06-13 12:21 | HST/WFC3 | 5.846 | 6.307 | 8.55 | −0.0646(27) | +0.0356(27) |
| 2017-06-14 07:25 | HST/WFC3 | 5.846 | 6.317 | 8.49 | +0.0932(22) | +0.0542(24) |
| 2017-12-09 14:36 | Keck II/NIRC2 | 5.910 | 5.875 | 9.58 | −0.1443(20) | +0.0102(24) |

Table notes:

[a.] The distance from the Sun to the target is $r$ and from the observer to the target is $\Delta$. The phase angle $g$ is the angular separation between the observer and Sun as seen from the target.

[b.] Visit mean relative right ascension $\Delta x$ and relative declination $\Delta y$ are computed as $\Delta x = (\alpha_2 - \alpha_1)\cos(\delta_1)$ and $\Delta y = \delta_2 - \delta_1$, where $\alpha$ is right ascension, $\delta$ is declination, and subscripts 1 and 2 refer to Patroclus and Menoetius, respectively. In parentheses are estimates of ±1-$\sigma$ astrometric uncertainties in the final 2 digits.

Our orbit fitting procedures have been described in prior publications and we refer the reader to them for additional details (e.g., Grundy et al. 2011; 2015). We used the Amoeba algorithm to find values of the Keplerian orbital elements that minimize residuals between observed and predicted relative astrometry, accounting for the heliocentric motions of Earth and Patroclus. We repeated the orbit-fitting exercise 10,000 times, each for a different realization of the MCMC sampling of the astrometric PDF, using a preliminary fit to the mean astrometry to set the initial simplex for each of the minimizations. The result was a cloud of solutions sampling the PDF in orbital element space. For Kuiper belt binaries, it is often the case that there are two possible solutions, mirror images of one another through the plane of the sky. The much more rapid heliocentric motion of the Patroclus system provides sufficient parallax for us to exclude the mirror solution with high confidence ($\chi^2 > 10^4$). The orbital solution we report (with $\chi^2 = 5.9$) is the median of the PDF for each orbital element, with the standard deviation as its uncertainty (see Table 2). This solution has a near zero eccentricity of 0.0043 ± 0.0049. We are skeptical that the eccentricity is not zero, since a system with such a small separation relative to the sizes of the bodies should rapidly circularize its orbit through tidal interaction and dissipation. Small, marginally significant non-zero eccentricities often show up in orbit fits, especially when relatively few epochs of astrometry are available, since the fitting algorithm will exploit whatever free parameters are available to it to address noise in the astrometric measurements. Accordingly, we also performed a solution with the eccentricity forced to zero. Although the goodness of fit of the forced-circular solution is slightly worse ($\chi^2 = 6.4$), we are adopting it as our preferred solution, since we do not have enough data to convince ourselves that the non-zero eccentricity is real.

| Table 2 - Orbit Solutions and 1-$\sigma$ Uncertainties for Patroclus and Menoetius | | | |
|---|---|---|---|
| Parameter | | Circular orbit | Unrestricted orbit |
| **Fitted elements**[a] | | $X^2 = 6.4$ | $X^2 = 5.9$ |
| Period (days) | $P$ | 4.282680 ± 0.000063 | 4.282696 ± 0.000074 |
| Semimajor axis (km) | $a$ | 688.5 ± 4.7 | 688.4 ± 4.7 |
| Eccentricity | $e$ | 0 (fixed) | 0.0043 ± 0.0049 |
| Inclination[b] (deg) | $i$ | 164.11 ± 0.61 | 164.13 ± 0.61 |
| Mean longitude[b] at epoch[c] (deg) | $\epsilon$ | 110.1 ± 3.6 | 109.7 ± 3.4 |
| Longitude of asc. node[b] (deg) | $\Omega$ | 269.3 ± 1.8 | 269.0 ± 1.7 |
| Longitude of periapsis[b] (deg) | $\varpi$ | N/A | 47 ± 76 |
| **Derived parameters** | | | |
| Standard gravitational parameter $GM_{sys}$ (km$^3$ s$^{-2}$) | $\mu$ | 0.0941 ± 0.0019 | 0.0940 ± 0.0019 |
| System mass (10$^{18}$ kg) | $M_{sys}$ | 1.410 ± 0.029 | 1.409 ± 0.029 |
| Orbit pole right ascension[b] (deg) | $\alpha_{pole}$ | 179.3 ± 1.8 | 179.0 ± 1.6 |
| Orbit pole declination[b] (deg) | $\delta_{pole}$ | −74.11 ± 0.60 | −74.13 ± 0.59 |
| Orbit pole ecliptic longitude[d] (deg) | $\lambda_{pole}$ | 234.2 ± 1.2 | 234.2 ± 1.1 |
| Orbit pole ecliptic latitude[d] (deg) | $\beta_{pole}$ | −62.11 ± 0.53 | −62.18 ± 0.52 |

Table notes:

[a.] Elements are for Menoetius relative to Patroclus. The average sky plane residual for the circular orbit solution is 2.1 mas and the maximum is 4.2 mas; $\chi^2$ is 6.4, based on the 7 epochs of observations. For the unrestricted orbit solution the average residual is 2.0 mas, and the maximum is 3.9 mas.

[b.] Referenced to J2000 equatorial frame.

[c.] The epoch is Julian date 2457900, corresponding to 2017 May 26 12:00 UTC.

[d.] Referenced to J2000 ecliptic frame.

Our new orbit solution can be compared to the Marchis et al. (2006) orbit solution, which was based on a completely independent set of observations obtained during 2004-2005. They reported a period of 4.283 ± 0.004 days and semimajor axis of 680 ± 20 km, with an orbit pole located at $\lambda$ = 236 ± 5 degrees, $\beta$ = −61 ± 1 degrees (J2000 ecliptic longitude and latitude). Our preferred circular orbit with parameters $P$ = 4.282680 ± 0.000063 days, $a$ = 688.5 ± 4.7 km, $\lambda$ = 234.2 ± 1.2 degrees, and $\beta$ = −62.11 ± 0.53 degrees is statistically consistent with the Marchis et al. orbit solution, albeit with smaller uncertainties. We can also compare our orbital period with the lightcurve period of 4.3125 ± 0.0125 reported by Oey (2012). The two periods are compatible at the 2.4 sigma level, consistent with the spins of Patroclus and Menoetius being tidally synchronized to the orbital period.

## Mutual event predictions

We predicted mutual event times from our circular orbit solution via a simple model that collapses the problem to a two-dimensional search for intersections between circles on the instantaneous sky plane as seen from Earth. A circle of radius 49 km, representing Patroclus, is fixed at the origin. A slightly smaller circle of radius 45 km representing Menoetius is placed at the satellite's position relative to Patroclus as seen from Earth, based on our orbit solution. These sizes are the polar radii from Buie et al. (2015), chosen because mutual events in a tidally-locked system would be viewed when the long-axes of prolate bodies are oriented toward the observer and/or Sun. This is a conservative assumption in the sense that the *b* axes of the bodies are larger than their *c* axes, so by assuming circular limb profiles at the polar radii, we predict minimum event durations. We test for intersections between these circles at one minute intervals to obtain the start, middle, and end times for occultation events. For whichever object is closer to the Sun, a third, appropriately-sized circle is placed to represent its shadow at the distance of the background body. Intersections between this circle and that representing the background body provide times for the eclipse events. For both occultation and eclipse components of events, the start, middle, and end times for an observer located at Earth are provided at http://www2.lowell.edu/~grundy/abstracts/2018.Patroclus.html, along with an indication of the event depth. This is a value between zero and one, with a value near zero indicating a grazing event and a value of one indicating perfect alignment of the centers of the relevant circles. The number is calculated as $S_{min}/(r_1+r_2)$ where $S_{min}$ is the minimum separation of the centers and $r_1$ and $r_2$ are the 49 and 45 km assumed projected radii of Patroclus and Menoetius, respectively. To translate these numbers into lightcurve predictions requires assumption of a photometric function to account for the center-to-limb brightness distribution. We are able to provide such predictions on request, but they are outside the scope of this note. We predict the mutual event season to begin in October 2017 with occultation events running through January 2018. There is then a brief hiatus before eclipse events start in February, turning into eclipse plus occultation events in June. Although the Earth is mostly on the wrong side of the Sun to observe them during the remainder of 2018, occultation plus eclipse events continue into 2019. The eclipse events extend through March 2019, while the occultation events end in June 2019. Patroclus is in opposition in early March 2018 and also in early April 2019. Events near opposition are especially convenient for Earth-based observers, since the target is above the horizon for more hours of the night. Times near new moon are also favorable, as the sky is darker. New moons near opposition include 2018 February 15, 2018 March 17, 2019 March 6, and 2019 April 5. The deepest events as measured by $S_{min}/(r_1+r_2)$ occur during July to October of 2018, although the system is poorly placed for Earth-based observations then.

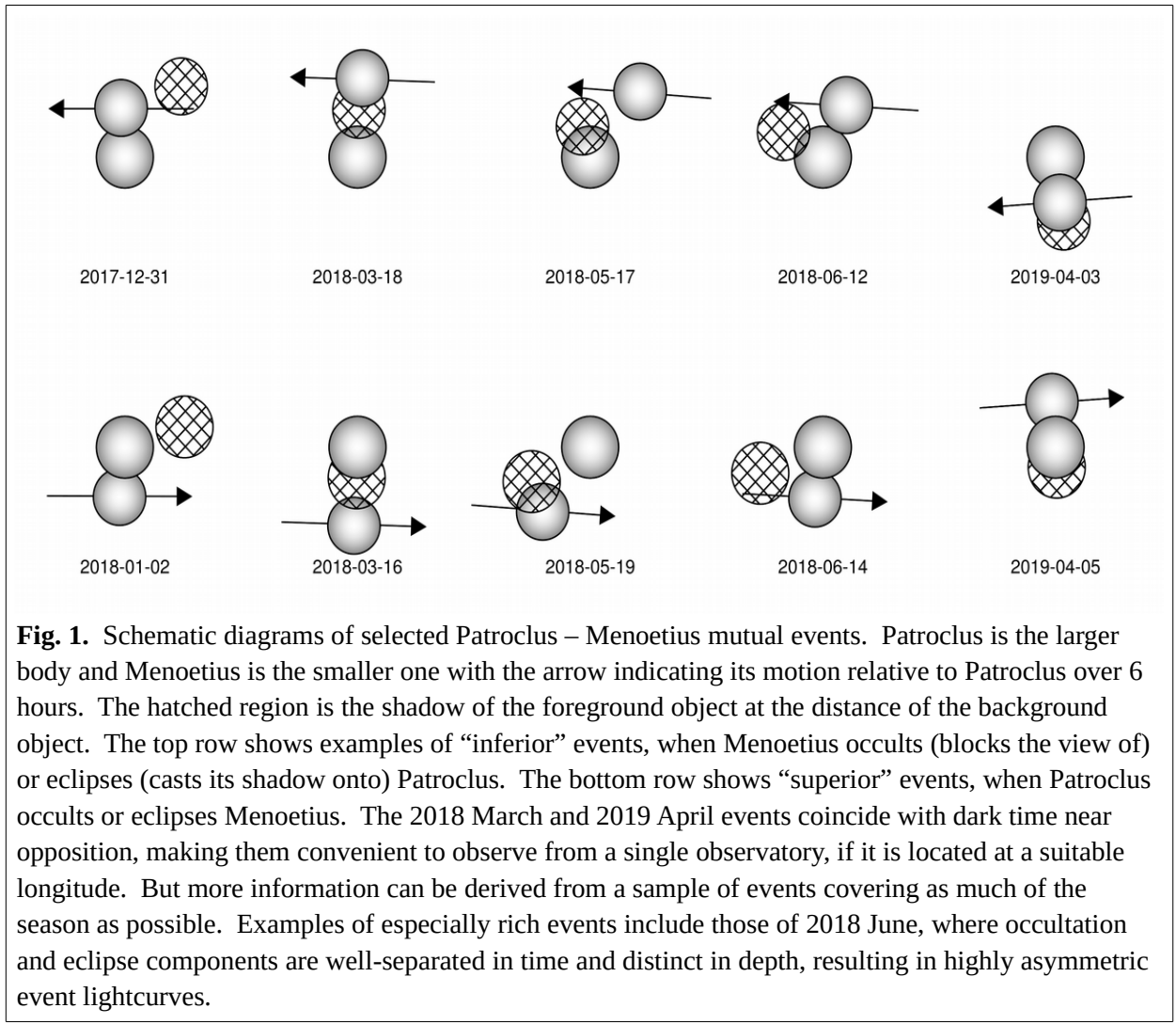

**Fig. 1.** Schematic diagrams of selected Patroclus – Menoetius mutual events. Patroclus is the larger body and Menoetius is the smaller one with the arrow indicating its motion relative to Patroclus over 6 hours. The hatched region is the shadow of the foreground object at the distance of the background object. The top row shows examples of "inferior" events, when Menoetius occults (blocks the view of) or eclipses (casts its shadow onto) Patroclus. The bottom row shows "superior" events, when Patroclus occults or eclipses Menoetius. The 2018 March and 2019 April events coincide with dark time near opposition, making them convenient to observe from a single observatory, if it is located at a suitable longitude. But more information can be derived from a sample of events covering as much of the season as possible. Examples of especially rich events include those of 2018 June, where occultation and eclipse components are well-separated in time and distinct in depth, resulting in highly asymmetric event lightcurves.

A number of factors add uncertainty to our mutual event predictions. The uncertainty in the orbital longitude (±3.6°), or 1% of the orbit, translates to a timing uncertainty for events of ±1 hour. Adding to this, the uncertainty in the orbital period leads to an additional contribution to the orbital longitude uncertainty that grows monotonically over time. Our estimate of the period is 4.282680 ± 0.000063 days, corresponding to a 1-σ uncertainty of roughly ±1 part in 68,000. After $n$ orbital periods have passed, this contribution to the longitude uncertainty will have grown to ±$n$ parts in 68,000, and so on. Approximately 28 orbital periods elapsed between the end of our HST observations and the start of the mutual event season in October 2017, corresponding to a longitude uncertainty of 1/2,400 of an orbit, or a timing uncertainty of ±2.5 minutes. By the end of the mutual event season in June 2019, that timing uncertainty will have grown to ±16 minutes, still small compared to the ±1 hour timing uncertainty from the uncertainty in orbital longitude. The uncertainty could be collapsed by incorporation of constraints from successful mutual event observations into the orbit solution, and we anticipate doing that as soon as suitable data become available[1]. Additional uncertainties come from the orientation of the orbit, and from the sizes and shapes of the objects. Tilting the orbit one way or the other can shift the timing of the overall mutual event season, and can modify the depths and durations of individual events, but has relatively little effect on their mid-times. Shrinking the sizes of the bodies reduces the duration of events, and also the duration of the mutual event season. Deviations from the circular limb profiles that we have assumed could have similar effects. Buie et al. (2015) estimated the *b* axis radii as 59 and 54 km, for Patroclus and Menoetius, respectively, slightly larger than their 49 and 45 km *c* axis radii that we used for our circular profiles. The larger *b* axis sizes imply somewhat longer events.

## Summary

We report new, independently-determined Keplerian orbit solutions for the Patroclus – Menoetius binary system that is consistent with the Marchis et al. (2006) orbit solution, but with smaller uncertainties. We use our new orbit to predict timing of mutual events observable from Earth during the coming mutual event season that runs from fall 2017 through summer 2019. Observations of the mutual events can be exploited to learn more about the system ahead of its exploration by the Lucy mission in 2033. Even just a single well-sampled event lightcurve can reduce the uncertainty in orbital longitude. Comparison of the timing of superior and inferior events can be used to test for non-zero orbital eccentricity. With a large number of event lightcurves including events both early and late in the mutual event season, it becomes possible to simultaneously constrain the orbital orientation and the bodies' limb profiles as seen along the direction of the tidal axis between the two bodies. Other goals that can be addressed using mutual event observations are to study time-dependent thermal behavior as done by Mueller et al. (2010), to constrain limb darkening photometric behavior, and to search for and localize albedo contrasts on the inward-facing hemispheres of the two objects. They can also be used to search for differences in photometric or spectral properties between the two bodies. At http://www2.lowell.edu/~grundy/abstracts/2018.Patroclus.html we provide a table of event times suitable for planning observations during the upcoming mutual event season. Updates will be provided there as they become available, and we encourage observers to collaborate in pursuit of these opportunities.

## Acknowledgments

This work is based in part on NASA/ESA Hubble Space Telescope program 14928. Support for this program was provided by NASA through a grant from the Space Telescope Science Institute (STScI), operated by the Association of Universities for Research in Astronomy, Inc., under NASA contract NAS 5-26555. Additional data were obtained at the W.M. Keck

---

1  We have just received word of the successful detection of two events by N. Pinilla-Alonso, E. Fernandez Valenzuela, R. Duffard, and J. Licandro. They observed the start of the 2017 Nov. 23 event and the full 2017 Dec. 08 event. The latter occurred roughly ¾ hour later than our predicted time, suggesting other events could be similarly delayed (Pinilla-Alonso and collaborators personal communication).

Observatory, a scientific partnership of the California Institute of Technology, the University of California, and NASA made possible by the generous financial support of the W.M. Keck Foundation. These data were obtained from telescope time allocated to NASA through the agency's scientific partnership with the California Institute of Technology and the University of California and their acquisition was supported by NASA Keck PI Data Awards, administered by the NASA Exoplanet Science Institute.  The authors wish to recognize and acknowledge the significant cultural role and reverence of the summit of Mauna Kea within the indigenous Hawaiian community.  We are grateful to have been able to observe from this mountain.  We also thank the free and open source software communities for empowering us with key tools used to complete this project, notably Linux, the GNU tools, LibreOffice, MySQL, the Astronomy User's Library, Evolution, Python, and FVWM.